\documentclass[%
 preprint,
 tightenlines,
 superscriptaddress,
 amsmath,amssymb,
 aps,
pra,
]{revtex4-1}

\usepackage{multibib}
\usepackage{soul}
\usepackage[colorlinks=true,allcolors=blue]{hyperref}
\usepackage{cleveref}
\crefname{equation}{equation}{equations}
\Crefname{equation}{Equation}{Equations}% For beginning \Cref
\crefrangelabelformat{equation}{(#3#1#4--#5#2#6)}
\crefmultiformat{equation}{equations (#2#1#3}{, #2#1#3)}{#2#1#3}{#2#1#3}
\Crefmultiformat{equation}{Equations (#2#1#3}{, #2#1#3)}{#2#1#3}{#2#1#3}

\usepackage{multirow}
\usepackage[normalem]{ulem}
\usepackage{color}
\usepackage{subfigure}
\usepackage{graphicx}
\usepackage{epstopdf}
\usepackage{dcolumn}% Align table columns on decimal point
\usepackage{bm}% bold math
\usepackage[]{lineno}% Enable numbering of text and display math
\newcommand*\phantomsubfigure[1]{\subfigure{\label{#1}}}

\begin{document}
\title{Intrinsic optimization using stochastic nanomagnets}

\author{Brian Sutton}
\email{bmsutton@purdue.edu}
\affiliation{School of Electrical and Computer Engineering, Purdue University, West Lafayette, IN, 47907}
\author{Kerem Yunus Camsari}
\affiliation{School of Electrical and Computer Engineering, Purdue University, West Lafayette, IN, 47907}
\author{Behtash Behin-Aein}
\affiliation{GLOBALFOUNDRIES Inc. USA, Santa Clara, CA 95054}
\author{Supriyo Datta}
\email{datta@purdue.edu}
\affiliation{School of Electrical and Computer Engineering, Purdue University, West Lafayette, IN, 47907}

\date{March 25, 2017}

\begin{abstract}
This paper draws attention to a hardware system which can be engineered so that its intrinsic physics is described by the generalized Ising model and can encode the solution to many important NP-hard problems as its ground state. The basic constituents are stochastic nanomagnets which switch randomly between the $\pm 1$ Ising states and can be monitored continuously with standard electronics. Their mutual interactions can be short or long range, and their strengths can be reconfigured as needed to solve specific problems and to anneal the system at room temperature. The natural laws of statistical mechanics guide the network of stochastic nanomagnets at GHz speeds through the collective states with an emphasis on the low energy states that represent optimal solutions. As proof-of-concept, we present simulation results for standard NP-complete examples including a 16-city traveling salesman problem using experimentally benchmarked models for spin-transfer torque driven stochastic nanomagnets.
\end{abstract}

\maketitle

%TC:break intro
\section*{\label{sec:intro}Introduction}

The use of Ising computers to solve NP-hard problems has a rich heritage in both theory\cite{barahona_computational_1982} and practice. These computers seek to solve a wide range of optimization problems by encoding the solution to the problem as the ground-state of an Ising energy expression. Many diverse systems have been proposed to solve NP-hard optimization problems such as those based on simulated annealing \cite{kirkpatrick_optimization_1983}, DNA\cite{adleman_molecular_1994, ouyang_dna_1997}, quantum annealing \cite{johnson_quantum_2011,perdomo-ortiz_finding_2012}, Cellular Neural Networks  \cite{chua_cellular_1988,chua_cnn_1993,ercsey-ravasz_cellular_2009}, CMOS \cite{yamaoka_20k-spin_2016}, trapped ions \cite{kim_quantum_2010}, electromechanics \cite{mahboob_electromechanical_2016}, optics  \cite{shaked_optical_2007,oltean_solving_2008,utsunomiya_mapping_2011,wu_optical_2014,wang_coherent_2013,marandi_network_2014,mcmahon_fully-programmable_2016,inagaki_coherent_2016}, and magnets \cite{bhanja_non-boolean_2015,arnalds_new_2016,behin-aein_building_2016}. A common objective of many of the Ising-based approaches is the identification of hardware configurations that can efficiently solve optimization problems of interest.

In this letter, we demonstrate the possibility of a hardware implementation that does not just mimic the Ising model, but embodies it as a part of its natural physics\cite{bhanja_non-boolean_2015,arnalds_new_2016,behin-aein_building_2016}. It uses a network of $N$ ``soft'' nanomagnets operating in a stochastic manner\cite{locatelli_spin-torque_2014}, each with an energy barrier $\Delta$ comparable to $k_{\text B}T$ so that they switch between the two Ising states, $\pm 1$, on time scales $\tau \sim \tau_0 \exp(\Delta/k_{\text B}T)$  where $\tau_0 \sim 0.1-1$ ns. The natural laws of statistical mechanics guide the network through the $2^N$ collective states at GHz rates, with an emphasis on low energy states. We show how an optimization problem of interest is solved by engineering the spin-mediated magnet-magnet interactions to encode the problem solution and to simulate annealing without any change in temperature simply by continuously adjusting their overall strength. As proof-of-concept for the potential applications of this natural Ising computer, we present detailed simulation results for standard NP-complete examples, including a 16-city traveling salesman problem. This involves using experimentally benchmarked modules to simulate a suitably designed network of 225 stochastic nanomagnets and letting the hardware itself rapidly identify solutions within the $2^{225}$  possibilities. It should be possible to integrate such hardware into standard solid state circuits, which will govern the scalability of the solution.

%TC:break intro2
The Ising Hamiltonian for a collection of spins, $S_i$, which can take on one of two values, $\pm 1$, 
\begin{equation}
\label{eq:ising} H = -\sum_{i,j} J_{ij} S_i S_j - \sum_i h_i S_i
\end{equation}
was originally developed to describe ferromagnetism where the $J_{ij}$ are positive numbers representing an exchange interaction between neighboring spins $S_i$ and $S_j$, while $h_i$ represents a local magnetic field for spin $S_i$. Classically, different spin configurations $\sigma\{S_i\}$ have a probability proportional to $\exp(-H(\sigma)/k_{\rm B}T)$, $T$ being the temperature, and $k_{\text B}$, the Boltzmann constant. At low temperatures, the system should be in its ground state $\sigma_\text{G}$, the state with the lowest energy $H(\sigma)$. With $h_i = 0$, and positive $J_{ij}$, it is easy to see that the ground state is the ferromagnetic configuration $\sigma_{\text F}$ with all spins parallel.

Much of the interest in the Ising Hamiltonian arises from the demonstration of many direct mappings of NP-complete and NP-hard problems to the model \cite{barahona_computational_1982,lucas_ising_2014,de_las_cuevas_simple_2016} such that the desired solution is represented by the spin configuration $\sigma$ corresponding to the ground state. However, in general this mapping may require a large number of spins, and may require the parameters $J_{ij}$ and $h_i$ to take on a wide range of values, both positive and negative. Finding the ground state of this artificial spin glass is the essence of Ising computing, and broadly speaking it involves abstractly representing an array of spins, their coupling, and thermal noise through software and hardware that attempts to harness the efficiencies of physical equivalence{\cite{khasanvis_physically_2015}. These representations may take the form of abstract models of the spins, the use of random number generators to produce noise, and logical or digital adders for the weighted summing. If enough layers of abstraction can be eliminated, the underlying hardware will inherently solve a given problem as part of its natural, intrinsic operation and this should be reflected in increased speed and efficiency.

\section*{\label{sec:exchange}Engineering Correlations Through Spin Currents}

Here we describe a natural hardware for an Ising computer based on the representation of an Ising spin $S_k$ by the magnetization $m$ of a stochastic nanomagnet(SNM), which we believe will compare well with other alternative representations. These SNMs are in the ``telegraphic'' switching regime\cite{locatelli_spin-torque_2014,bapna_magnetostatic_2016} requiring the existence of a small barrier in the magnetic energy ($\Delta \approx k_{\rm B}T $), that gives a small, but definite preference for a given axis, with two preferred states $\pm 1$. In the absence of currents, these SNMs continually switch between $+1$ and $-1$ on the order of nanoseconds, and can be physically realized by a reduction of the magnetic grain volume\cite{locatelli_noise-enhanced_2014} or by designing weak perpendicular magnetic anisotropy (PMA) magnets\cite{cowburn1999single}. Figure \ref{fig:fig1} shows the response of such a monodomain PMA magnet in the presence of an external spin current in the direction of the magnet's easy axis.

How do we couple the SNMs to implement the Ising Hamiltonian of Equation \eqref{eq:ising}? The usual forms of coupling involve dipolar or exchange interactions that are too limited in range and weightability. Instead, one possibility is an architecture\cite{behin-aein_building_2016} that uses charge currents which can be readily converted locally into spin currents through the spin Hall effect (SHE). These charge currents can be arbitrarily long-range and the total number of cross-couplings is only limited by considerations of routing congestion and delay. The couplings may also be confined to nearest-neighbors, simplifying the hardware design complexity while promoting scalability and retaining universality\cite{de_las_cuevas_simple_2016}. 

The Ising Hamiltonian of Equation \eqref{eq:ising} can be implemented by exposing each SNM $m_k$ to a spin current $I_k$
\begin{equation}
\label{eq:per_magnet_current} I_k(\{m_j\}) = \frac{2q\alpha}{\hbar} \left(h_k + \sum_j 2J_{kj} m_j\right)
\end{equation}
which has a constant bias determined by $h_k$ together with a term proportional to the magnetization of the $j^{\text{th}}$ SNM $m_j$. The future state of magnet $m_k$ at time $(t+\Delta t)$ is related to the state of the other magnets at time $t$ through the current $I_k$. This expression is derived analytically in the following section using the Fokker-Planck equation for the system\cite{butler_switching_2012}.

The spin current $I_k$ can be generated using well-established phenomena and the prospects for physical realization of such a system are discussed later in this paper. The distinguishing feature of the present proposal arises from the intrinsic stochasticity of SNMs and their biasing through the use of weighted spin currents (Figure \ref{fig:fig1a}). How the SNMs are interconnected to implement Equation \eqref{eq:per_magnet_current} can evolve as the field progresses.

Getting a large system to reach its true ground state is non-trivial as it tends to get stuck in local minima\cite{aaronson_guest_2005}. It is common to guide the system towards the ground state through a process of ``annealing''\cite{kirkpatrick_optimization_1983} which is carried out differently in different hardware implementations. For example, systems based on superconducting flux qubits make use of quantum tunneling, which is referred to as quantum annealing\cite{kadowaki_quantum_1998}, whereas classical CMOS approaches make use of random number generators\cite{cheemalavagu_probabilistic_2005} to produce random transitions out of local minima.

For our system of coupled SNMs, random noise is naturally present and can be easily controlled (Figure \ref{fig:fig1a}}), causing the system of SNMs coupled according to Equation \eqref{eq:per_magnet_current} to explore the configuration space of the problem on a nanosecond timescale. Annealing could be performed through a controlled lowering of the actual temperature, or equivalently through a controlled increase in the magnitude of the current $I_k$, even at room temperature. It has been noted that certain annealing schedules can guarantee convergence to the true ground state, but these schedules may be too slow to be used in practice\cite{geman_stochastic_1984}. This paper only presents a straightforward annealing process and does not seek out optimal annealing schedules. Consequently, as we show in one of our combinatorial optimization examples, we may find only an approximate solution which, however, may be adequate for many practical problems.

\subsection*{\label{sec:fokker}Steady-State Fokker-Planck Description}

Our goal is to interconnect magnets such that their equilibrium state is governed by Boltzmann statistics with thermal noise as an inherent characteristic of the system. To see that this is possible, consider a system of $N$ magnets where we want 
\begin{equation}
\label{eq:app_a1}\rho(m_1,\cdots, m_N) = \rho_0 e^{-E(m_1,\cdots, m_N)/k_B T}
\end{equation}
and
\begin{equation}
\label{eq:app_a13} E(m_1,\cdots, m_N) = \sum_i \left(A_i m_i^2 + h_i m_i\right) + \sum_{i,j} J_{ij}m_i m_j
\end{equation}
where $m_k$ represents the z-component of the magnets.

Suppose each magnet is driven by a spin current derived from the others. We start with the Fokker-Planck equation \cite{butler_switching_2012} for the $N$-magnet system:
\begin{align}
\label{eq:app_a3} \frac{\partial \rho}{\partial \tau} &= \frac{\partial}{\partial m_k} \left\{(1-m_k^2)\left[(i_k - m_k)\rho + \frac{1}{2\Delta_k}\frac{\partial \rho}{\partial m_k}\right] \right\}
\end{align}
where $\Delta_k = \mu_0 H_k M_s V / 2k_\text{B} T$ and $i_k = I_k/I_0$ with $I_0$ as the critical switching spin current $I_0 = (2q\alpha/\hbar) 2\Delta_k k_\text{B}T$. At equilibrium, $\partial \rho/ \partial \tau = 0$ yielding from \eqref{eq:app_a1} and \eqref{eq:app_a3}:
\begin{equation}
\label{eq:app_a14} \frac{\partial (\ln \rho)}{\partial m_k} = - 2 \Delta_k (i_k-m_k)
\end{equation}
\begin{equation}
\label{eq:app_a15} \frac{\partial (\ln \rho)}{\partial m_k} = -\frac{1}{k_B T} \left(2A_km_k + h_k + \sum_j (J_{kj} + J_{jk})m_j\right)
\end{equation}
respectively. Comparing equations \eqref{eq:app_a14} and \eqref{eq:app_a15} while assuming symmetric coupling, $J_{kj} \equiv J_{jk}$, for the system we find
\begin{equation}
\Delta_k = -A_k/k_BT
\end{equation} 
and arrive at \eqref{eq:per_magnet_current}:
\begin{equation}
\notag i_k = \frac{h_k + \sum_j 2J_{kj}m_j}{\mu_0 H_K M_S V}
\end{equation}

\subsection*{\label{sec:modeling}Stochastic Landau-Lifshitz-Gilbert (LLG) Model}

In this section we briefly describe the simulation framework and stochastic LLG model used throughout this paper. We start with the LLG equation \cite{butler_switching_2012} for a monodomain magnet with magnetization $m_i$ in the presence of a spin current ($\vec{I}_s = I_0 \hat z)$
\begin{equation}
 (1+\alpha^2)\frac{d\hat m_i}{dt} = -|\gamma|{\hat m_i \times \vec{H}_i} - \alpha |\gamma| (\hat m_i \times \hat m_i \times \vec{H}_i)    +  \frac{1}{q  N_i}(\hat m_i \times \vec{I}_{Si} \times \hat m_i)  + \left(\frac{\alpha}{q N_i} (\hat m_i \times \vec{I}_{Si})\right)
\label{eq:llg}
\end{equation}
The magnetic thermal noise enters the equation through the effective field of the magnet, $H_i = H_0 + H_n$, as an uncorrelated external magnetic field in three dimensions with the following mean and variance:
\begin{equation}
\label{eq:app_b2} \langle {H_n^{\vec{r}}} \rangle = 0\text{, }  \quad \langle |H_n^{\vec{r}}|^2 \rangle =  \frac{2\alpha \rm kT}{|\gamma| \rm M_s Vol.}
\end{equation}
The numerical model is implemented as an equivalent circuit for SPICE-like simulators and reproduces the equilibrium (Boltzmann) distribution from a Fokker-Planck Equation \cite{butler_switching_2012}.

A given system of magnets is simulated using a collection of independent, though current-coupled, stochastic LLG models. Delays associated with the communication from one magnet to the next are neglected assuming that the response time of the nanomagnets is much greater than associated wire-delays. Presently, the attempt time $\tau$ of experimental nanomagnets is on the order of $\sim\mu$s to $\sim$ms \cite{koch_thermally_2000,locatelli_noise-enhanced_2014,bapna_magnetostatic_2016}. With additional scaling, the response times of these magnets will continue to improve \cite{urazhdin_current-driven_2003} and should approach the $\sim$ns times discussed in this paper. With response times $\sim$ns, our simulations show that even routing delays on the order of 100s of ps do not affect the results materially. Using nearest-neighbor Ising approaches or other constraining design decisions it should be possible to limit routing delays to shorter values. However, if the routing delay is comparable to the intrinsic response time of the nanomagnets then it would be important to include their effect in the simulation. 

Many options exist, please see the final section, for physical realization of the proposed system of stochastic nanomagnets. For the simulations in this paper we simply use Equation \eqref{eq:per_magnet_current} without assuming any specific hardware to implement it, since it is likely that better alternatives will emerge in the near future, given the rapid pace of discovery in the field of spintronics, see for example\cite{camsari2015modular,liu_spin-torque_2012,heron2014deterministic,sanchez2013spin}.

\section*{\label{sec:comb_opt}Combinatorial Optimization}

We will focus on two specific examples to demonstrate the ability of such an engineered spin glass to solve problems of interest\cite{karp_reducibility_1972}: an instructive example based on the satisfiability problem (SAT), and a representative example based on the traveling salesman problem (TSP). The first known NP-complete problem is the problem of Boolean satisfiability\cite{cook_complexity_1971}, namely, deciding if some assignment of boolean variables $\{x_i\}$ exists that satisfies a given conjunctive normal form (CNF) expression. Finding the collection of inputs that makes the clauses of the CNF expression true is computationally difficult, but easy to verify.

It is known that any given CNF expression can be mapped to a collection of Ising contraints using the fundamental building blocks of NOT ($m_1 = \bar{m_2}$), AND ($m_1 = m_2 \wedge m_3$), and OR ($m_1 = m_2 \vee m_3$) each subject to the Ising constraints given by \cite{bian_ising_2010}: 
\begin{align}
\label{eq:Hnot} H_\text{NOT} &= 1-(-m_1 m_2)\\
\label{eq:Hand} H_\text{AND} &= 3-(-m_2m_3 + 2m_1m_2 + 2m_1m_3) - (-2m_1 + m_2 + m_3) \\
\label{eq:Hor} H_\text{OR} &= 3-(-m_2m_3 + 2m_1m_2 + 2m_1m_3) - (2m_1 - m_2 - m_3) 
\end{align}
Using these building blocks, a network capable of finding the truth table for XOR ($m_1 = (m_2 \vee m_3) \wedge \overline{(m_2 \wedge m_3)}$) was prepared (Figure \ref{fig:fig2}). For simplicity, the solution uses a naive method to construct the network and leverages the use of ancillary spins to represent $(m_2 \vee m_3)$ and $(m_2 \wedge m_3)$ respectively (note that four spins could have been used \cite{biamonte_nonperturbative_2008}). The array of spins from Figure \ref{fig:fig2b} are connected as specified by \cref{eq:Hnot,eq:Hand,eq:Hor}, driven by a reference current $I_0$. As the magnets explore the configuration space, their outputs are digitized and used to compute the overall energy of the system (Figure \ref{fig:fig2c}). The regions of zero energy correspond to solutions of the problem. The digitized outputs are aggregated to determine their probability of occurrence. By looking at the first three bits of the most probable outputs, the solution to the problem can be directly found (Figures \ref{fig:fig2d} and \ref{fig:fig2e}). While this problem helps convey the essence of the approach, a more demonstrative application is worth considering.

%TC:break TSP
The decision form of the TSP is NP-complete, that is, for a collection of $N$ cities, does there exist a closed path for which each city is visited exactly once that has a tour length less than some value $d$? Finding tours that satisfy this problem is computationally challenging and also of great practical interest. There are well-known mappings that translate the TSP to the Ising model \cite{schneider_stochastic_2006,lucas_ising_2014}. Here we adopt the following:

\begin{align}
\label{eq:Htsp} H &= \sum_{v=1}^N\left(1-\sum_{j=1}^N x_{v,j}\right)^2 + \sum_{j=1}^N\left(1-\sum_{v=1}^N x_{v,j}\right)^2 + \lambda \sum_{uvj}W_{(uv)}x_{u,j}x_{v,j+1}
\end{align}
where $x_{i,j}$ is a Boolean variable that is TRUE when city $i$ is stop number $j$ and FALSE otherwise, and $W_{(uv)}$ are directed weights based on the distance between cities $u$ and $v$. This Hamiltonian is mapped to a spin system by replacing each $x_{ij}$ with $1/2 (m_{ij}+1)$ and weights $W_{(uv)}$ with $i_{(uv)}$ given by Equation \eqref{eq:per_magnet_current}.

If the interconnections between each city are symmetric, then a Boltzmann machine\cite{ackley_learning_1985} with each of the $2^{N\times N}$ states associated with an effective energy $H$ is realized, and the probability of the system visiting a particular state is proportional to $\exp(-H/k_{\rm B}T)$. In order to find low-energy, optimized states, direct annealing of the glass can be performed. Using the ulysses16 reference dataset\cite{reinelt_tsplib--_1991}, annealing of a problem specific magnetic array through control of the effective temperature was performed (Figure \ref{fig:fig3}). Two specific traits of interest arise, namely the energy decays in a sigmoidal relationship with the $\ln T$, and the specific heat of the system, $C(T) = (\langle E(T)^2\rangle - \langle E(T)\rangle^2)/k_{\rm B}T^2$, shows a defined peak about a critical temperature. At high temperatures, the system is disordered and corresponds to high energy states (Figure \ref{fig:fig3c}). As the temperature is reduced, the system continues to explore the energy landscape on a nanosecond timescale while gradually converging to a low-energy solution. For the given annealing profile and simulation duration, a low-energy, though not ideal, solution is found to the problem, highlighting the heuristic nature of the optimization\cite{schneider_stochastic_2006}. Note that in principle these simulation results could be obtained directly from actual hardware. For example, Figures \ref{fig:fig2d} and \ref{fig:fig3d} could be obtained by continuously monitoring the states of the individual SNMs using spin valves.

\section*{\label{sec:phys_rel}Considerations for Physical Realization}

Physical realization of these engineered spin glasses requires the integration of multiple functional elements as highlighted in Figure \ref{fig:fig2a}. The magnetization of each magnet $m_i$ is first sensed with a read unit. The signal produced by this read unit is then propagated to all of the magnets with couplings dependent on the read magnetization $m_i$. Each of these connections is independently weighted with weights $W_{(ij)}$ and provided as input, along with an on-site bias $B_i$ to the write units. The write units in turn influence and control the state of magnet $m_j$. 

There are a number of design options available for each functional unit as shown in Table \ref{tab:phys}. Write-control of the magnets can be affected through a number of means including the spin Hall effect (SHE)\cite{liu_spin-torque_2012} or perhaps through voltage control\cite{heron2014deterministic}. The use of the SHE effect provides a convenient mechanism with which to sum several, independently weighted, input currents. Readout of the magnetization can be accomplished using well-established tunnel junctions  \cite{parkin_magnetically_2003} which have been demonstrated for stochastic nanomagnets \cite{locatelli_noise-enhanced_2014}. Alternatively, readout could perhaps be accomplished using the inverse SHE \cite{liu_spin-torque_2012}. Assuming the use of a SHE material and tunnel junction stack, care must be given to accomodate the simultaneous use of write and read currents. One approach would be to introduce the use of a time-multiplexed scheme that disassociates the write and read operations\cite{sengupta_probabilistic_2016}. Alternatively, structures that provide write and read isolation may be used \cite{datta2012non}.

The ability to write and read the magnetization is of fundamental importance, however, once read, the likely weak signal must be amplified to satisfy the fanout requirements of the network. This transistor-like gain can be realized using all-spin based approaches \cite{datta2012non, behin-aein_building_2016} or perhaps with the use of a hybrid-CMOS design\cite{tangel_cmos_2004, sengupta_spin_2015}. These proposed approaches may introduce power dissipation challenges during the read operation, e.g. the short-circuit current produced with the use of amplifying inverters. Power dissipation considerations must be carefully evaluated to assess the viability of scaling the proposed system.

The output from the amplification stage can be selectively weighted so that a wide range of problems based on \eqref{eq:ising} can be encoded onto the network. The weighting of inputs can be based on an approach using re-programmable floating-gate voltages\cite{diep_spin_2014} that would enable the use of analog weights for the circuit. While floating-gate regulation would enable convenient re-programmability, the design would be complicated with the requirement for peripheral drivers to control the floating-gate array. Others proposals have suggested the use of memristors \cite{locatelli_spin-torque_2014,yang_memristive_2013, sengupta_spin_2015} or other programmable elements in a cross-bar like configuration \cite{sengupta_proposal_2016, sengupta_probabilistic_2016}, though with constrained fanout. Note that one weighting scheme that still retains the ability to encode NP-hard problems onto the network is with the use of $\{-1,0,1\}$ weights\cite{barahona_computational_1982}. Using this simple approach removes the necessity for tunable weights and instead relegates the problem to one of routing, connectivity, and area.

All of the simulations used in this paper assume a fully connected network of magnets in which each magnet talks to all other magnets. For small networks this is reasonable, however, such an assumption is invalid for large networks as the number of routes grows rapidly. Instead, different topologies\cite{bunyk_architectural_2014} and routing considerations must be made to account for congestion and long-distance communication. By limiting the connections to local-neighbors \cite{ercsey-ravasz_cellular_2009, yamaoka_20k-spin_2016}, the network may still be used to perform NP-hard optimization while also simplifying routing complexity. One design possibility is to leverage the lessons learned from the advances in the design of Field-Programmable Gate Array (FPGA) interconnects\cite{lemieux_design_2004}. FPGAs are designed with routing topologies that facilitate both short and long-range interconnections while also providing re-programmability.

The fidelity of the programmed weights and number of high-fanout signals needed for robust solutions may impose challenges on the selected weighting and routing schemes. Additionally, the propagation delay of these high-fanout signals must be balanced with the response time of the magnets in order for the system to be governed by \eqref{eq:ising}. While flexibility in the allowed weights and number of couplings is convenient for encoding problems onto the model\cite{lucas_ising_2014}, it is important to note that discrete nearest-neighbor couplings still retain NP-hardness\cite{barahona_computational_1982} and may greatly simplify the hardware design, improving scalability at the expense of increased encoding complexity and area. 

The main point of this paper is the remarkable high-speed search through Fock space enabled by the intrinsic physics of a network of stochastic nanomagnets interacting via spin-mediated interactions. We hope this work fosters an interest in the physical realization and exploration of stochastic nanomagnets as a viable Ising computer.

%\bibliographystyle{vancouver_bs}
%\bibliography{bib}% Produces the bibliography via BibTeX.

\begin{thebibliography}{10}

\bibitem{barahona_computational_1982}
Barahona, F.
\newblock On the computational complexity of {Ising} spin glass models.
\newblock {\em J. Phys. Math. Gen.} \textbf{15}, 3241--3253 (1982).

\bibitem{kirkpatrick_optimization_1983}
Kirkpatrick, S., Gelatt, C. D. \& Vecchi, M. P.
\newblock Optimization by simulated annealing.
\newblock {\em Science} \textbf{220}, 671--680 (1983).

\bibitem{adleman_molecular_1994}
Adleman, L. M.
\newblock Molecular computation of solutions to combinatorial problems.
\newblock {\em Science} \textbf{266}, 1021--1024 (1994).

\bibitem{ouyang_dna_1997}
Ouyang, Q., Kaplan, P.D., Liu, S. \& Libchaber, A.
\newblock DNA solution of the maximal clique problem.
\newblock {\em Science} \textbf{278}, 446--449 (1997).

\bibitem{johnson_quantum_2011}
Johnson, M. W. \emph{et al}. 
\newblock Quantum annealing with manufactured spins.
\newblock {\em Nature} \textbf{473}, 194--198 (2011).

\bibitem{perdomo-ortiz_finding_2012}
Perdomo-Ortiz, A., Dickson, N., Drew-Brook, M., Rose, G. \& Aspuru-Guzik, A.
\newblock Finding low-energy conformations of lattice protein models by quantum
  annealing.
\newblock {\em Sci. Rep.} \textbf{2}, 571 (2012).

\bibitem{chua_cellular_1988}
Chua, L. O. \& Yang L.
\newblock Cellular neural networks: applications.
\newblock {\em IEEE Trans. Circuits Syst.} \textbf{35}, 1273--1290 (1988).

\bibitem{chua_cnn_1993}
Chua, L. O. \& Roska, T.
\newblock The {CNN} paradigm.
\newblock {\em IEEE Trans. Circuits System. I, Fundam. Theory Appl.} \textbf{40}, 147--156 (1993).

\bibitem{ercsey-ravasz_cellular_2009}
Ercsey-Ravasz, M., Roska, T. \& N\'eda Z.
\newblock Cellular {Neural} {Networks} for {NP}-hard optimization.
\newblock {\em EURASIP J. Adv. Signal Process.}. \textbf{2009},
  2:1--2:7 (2009).

\bibitem{yamaoka_20k-spin_2016}
Yamaoka, M. \emph{et al}.
\newblock A 20k-spin ising chip to solve combinatorial optimization
  problems with CMOS annealing.
\newblock {\em IEEE J. Solid-State Circuits} \textbf{51}, 303--309 (2016).

\bibitem{kim_quantum_2010}
Kim, K. \emph{et al}.
\newblock Quantum simulation of frustrated {Ising} spins with trapped ions.
\newblock {\em Nature} \textbf{465}, 590--593 (2010).

\bibitem{mahboob_electromechanical_2016}
Mahboob, I., Okamoto, H. \& Yamaguchi, H.
\newblock An electromechanical {Ising} {Hamiltonian}.
\newblock {\em Sci. Adv.} \textbf{2}, e1600236 (2016).

\bibitem{shaked_optical_2007}
Shaked, N. T., Messika, S., Dolev, S. \& Rosen, J.
\newblock Optical solution for bounded {NP}-complete problems.
\newblock {\em Appl. Opt.} \textbf{46}, 711--724 (2007).

\bibitem{oltean_solving_2008}
Oltean, M.
\newblock Solving the {Hamiltonian} path problem with a light-based computer.
\newblock {\em Nat. Comput.}. \textbf{7}, 57--70 (2008).

\bibitem{utsunomiya_mapping_2011}
Utsunomiya, S., Takata, K. \& Yamamoto, Y.
\newblock Mapping of {Ising} models onto injection-locked laser systems.
\newblock {\em OOpt. Express} \textbf{19}, 18091--18108 (2011).

\bibitem{wu_optical_2014}
Wu, K., Garc\'ia~de Abajo, J., Soci, C., Ping~Shum, P. \& Zheludev, N.I.
\newblock An optical fiber network oracle for {NP}-complete problems.
\newblock {\em Light Sci. Appl.} \textbf{3}, e147 (2014).

\bibitem{wang_coherent_2013}
Wang, Z., Marandi, A., Wen, K., Byer, R.L. \& Yamamoto, Y.
\newblock Coherent {Ising} machine based on degenerate optical parametric oscillators.
\newblock {\em Phys. Rev. A.} \textbf{88}, 063853 (2013).

\bibitem{marandi_network_2014}
Marandi, A., Wang, Z., Takata, K., Byer, R. L. \& Yamamoto, Y.
\newblock Network of time-multiplexed optical parametric oscillators as a coherent {Ising} machine.
\newblock {\em Nat. Photon.} \textbf{8}, 937--942 (2014).

\bibitem{mcmahon_fully-programmable_2016}
McMahon, P. L. \emph{et al}.
\newblock A fully-programmable 100-spin coherent {Ising} machine with all-to-all connections.
\newblock {\em Science}. aah5178 (2016).

\bibitem{inagaki_coherent_2016}
Inagaki, T. \emph{et al}.
\newblock A coherent {Ising} machine for 2000-node optimization problems.
\newblock {\em Science} \textbf{354}, 603--606 (2016).


\bibitem{bhanja_non-boolean_2015}
Bhanja, S., Karunaratne, D. K., Panchumarthy, R., Rajaram, S. \& Sarkar, S.
\newblock Non-{Boolean} computing with nanomagnets for computer vision
  applications.
\newblock {\em Nature Nano.} \textbf{11}, 177--183 (2015).

\bibitem{arnalds_new_2016}
Arnalds, U.B. \emph{et al}.
\newblock A new look on the two-dimensional {Ising} model: thermal artificial
  spins.
\newblock {\em New J. Phys.} \textbf{18}, 023008 (2016).

\bibitem{behin-aein_building_2016}
Behin-Aein, B., Diep, V. \& Datta, S.
\newblock A building block for hardware belief networks.
\newblock {\em Sci. Rep.} \textbf{6} 29893 (2016).

\bibitem{locatelli_spin-torque_2014}
Locatelli, N., Cros, V. \& Grollier, J.
\newblock Spin-torque building blocks.
\newblock {\em Nature Mater.} \textbf{13}, 11--20 (2014).

\bibitem{lucas_ising_2014}
Lucas, A.
\newblock Ising formulations of many {NP} problems.
\newblock {\em Front. Physics.} \textbf{2}, 5 (2014).

\bibitem{de_las_cuevas_simple_2016}
De~las Cuevas, G. \& Cubitt, T. S.
\newblock Simple universal models capture all classical spin physics.
\newblock {\em Science} \textbf{351}, 1180--1183 (2016).

\bibitem{khasanvis_physically_2015}
Khasanvis, S. \emph{et al}.
\newblock Physically equivalent magneto-electric nanoarchitecture for
  probabilistic reasoning, Proceedings of the International Symposium
  on Nanoscale Architectures ({NANOARCH}), pp. 25--26 (2015).

\bibitem{bapna_magnetostatic_2016}
Bapna, M. \emph{et al}.
\newblock Magnetostatic effects on switching in small magnetic tunnel junctions.
\newblock {\em Appl. Phys. Lett.}. \textbf{108}, 022406 (2016).

\bibitem{locatelli_noise-enhanced_2014}
Locatelli, N. \emph{et al}.
\newblock Noise-enhanced synchronization of stochastic magnetic
  oscillators.
\newblock {\em Phys. Rev. Applied} \textbf{2}, 034009 (2014).

\bibitem{cowburn1999single}
Cowburn, R. P., Koltsov, D. K., Adeyeye, A. O., Welland, M. E. \& Tricker, D. M.
\newblock Single-domain circular nanomagnets.
\newblock {\em Phys. Rev. Lett.} \textbf{83}, 1042--1045 (1999).

\bibitem{butler_switching_2012}
Butler, W. H. \emph{et al}.
\newblock Switching distributions for perpendicular spin-torque devices within the macrospin approximation.
\newblock {\em IEEE Trans. Magn.} \textbf{48}, 4684--4700 (2012).



\bibitem{aaronson_guest_2005}
Aaronson, S.
\newblock Guest column: {NP}-complete problems and physical reality.
\newblock {\em SIGACT News} \textbf{36}, 30--52 (2005).

\bibitem{kadowaki_quantum_1998}
Kadowaki, T., Nishimori, H.
\newblock Quantum annealing in the transverse {Ising} model.
\newblock {\em Phys. Rev. E} \textbf{58}, 5355--5363 (1998).

\bibitem{cheemalavagu_probabilistic_2005}
Cheemalavagu, S., Korkmaz, P., Palem, K. V., Akgul. B. E. S. \& Chakrapani, L. N.
\newblock A probabilistic {CMOS} switch and its realization by exploiting
  noise, {Proceedings} of the {IFIP} international conference on very large scale integration, (2005).

\bibitem{geman_stochastic_1984}
Geman, S. \& Geman, D.
\newblock Stochastic relaxation, Gibbs distributions, and the Bayesian
  restoration of images.
\newblock {\em IEEE Trans. Pattern Anal. Mach. Intell.} \textbf{6}, 721--741 (1984).


\bibitem{koch_thermally_2000}
Koch, R. H. \emph{et al}.
\newblock Thermally assisted magnetization reversal in submicron-sized magnetic thin films.
\newblock {\em Phys. Rev. Lett.} \textbf{84}, 23 (2000).


\bibitem{urazhdin_current-driven_2003}
Urazhdin, S., Birge, N. O., Pratt, W. P. \& Bass, J.
\newblock Current-driven magnetic excitations in permalloy-based multilayer nanopillars.
\newblock {\em Phys. Rev. Lett.} \textbf{91}, 14 (2003).


\bibitem{camsari2015modular}
Camsari, K. Y., Ganguly, S. \& Datta, S.
\newblock Modular approach to spintronics.
\newblock {\em Sci. Rep.} \textbf{5}, 10571 (2015).


\bibitem{liu_spin-torque_2012}
Liu, L. \emph{et al}.
\newblock Spin-torque switching with the giant spin Hall effect of
  tantalum.
\newblock {\em Science} \textbf{336}, 555--558 (2012).

\bibitem{heron2014deterministic}
Heron, J. T. \emph{et al}.
\newblock Deterministic switching of ferromagnetism at room temperature using
  an electric field.
\newblock {\em Nature} \textbf{516}, 370--373 (2014).

\bibitem{sanchez2013spin}
Rojas S{\'a}nchez J. C. \emph{et al}.
\newblock Spin-to-charge conversion using Rashba coupling at the interface
  between non-magnetic materials.
\newblock {\em Nature Commun.} \textbf{4}, 2944 (2013).

\bibitem{karp_reducibility_1972}
Karp, R. M.
\newblock Reducibility among combinatorial problems, in \emph{Complexity of
  Computer Computations} (eds. Miller, R. E. \& Thatcher, J. W.) pp. 85--103 (Plenum Press, New York, 1972).

\bibitem{cook_complexity_1971}
Cook, S. A.
\newblock The complexity of theorem-proving procedures, in \emph{Proc. 3rd Ann. Symp. on Theory of Computing} 151-158 (ACM, 1971).

\bibitem{bian_ising_2010}
Bian, Z., Chudak, F., Macready, W. G. \& Rose, G.
\newblock The {Ising} model: teaching an old problem new tricks.
\newblock {\em D-Wave Systems} \textbf{2} (2010).

\bibitem{biamonte_nonperturbative_2008}
Biamonte, J. D.
\newblock Nonperturbativ $k$-body to two-body commuting conversion Hamiltonians and embedding problem instances into Ising spins.
\newblock {\em Phys. Rev. A} \textbf{77}, 052331 (2008).

\bibitem{schneider_stochastic_2006}
Schneider, J. J. \& Kirkpatrick, S.
\newblock \emph{Stochastic Optimization} (Springer, 2006).

\bibitem{ackley_learning_1985}
Ackley, D. H., Hinton, G. E. \& Sejnowski, T. J.
\newblock A learning algorithm for Boltzmann machines.
\newblock \emph{Cognitive Sci.} \textbf{9}, 147--169 (1985).

\bibitem{reinelt_tsplib--_1991}
Reinelt G.
\newblock {TSPLIB}--A traveling salesman problem library.
\newblock {\em ORSA Journal on Computing} \textbf{3}, 376 (1991).

\bibitem{parkin_magnetically_2003}
Parkin, S. \emph{et al}.
\newblock Magnetically engineered spintronic sensors and memory.
\newblock {\em Proceedings of the IEEE}. \textbf{91}, 661--680 (2003).

\bibitem{sengupta_probabilistic_2016}
Sengupta, A., Parsa, M., Han, B. \& Roy, K.
\newblock Probabilistic deep spiking neural systems enabled by magnetic tunnel junction.
\newblock {\em IEEE Trans. Electron Dev.} \textbf{63}, 2963--2970 (2016).

\bibitem{datta2012non}
Datta, S., Salahuddin, S. \& Behin-Aein, B.
\newblock Non-volatile spin switch for Boolean and non-Boolean logic.
\newblock {\em Appl. Phys. Lett.} \textbf{101}, 252411 (2012).

\bibitem{tangel_cmos_2004}
Tangel, A., \& Choi, K.
\newblock ``{The} {CMOS} {Inverter}'' as a comparator in ADC designs.
\newblock {\em Analog Integr. Circuits Signal Process.} \textbf{39}, 147--155 (2004).

\bibitem{sengupta_spin_2015}
Sengupta, A., Choday, S.H., Kim, Y. \& Roy, K.
\newblock Spin orbit torque based electronic neuron.
\newblock {\em App. Phys. Lett.}. \textbf{106}, 143701 (2015).

\bibitem{diep_spin_2014}
Diep, V. Q., Sutton, B., Behin-Aein, B. \& Datta, S.
\newblock Spin switches for compact implementation of neuron and synapse.
\newblock {\em Appl. Phys. Lett.}. \textbf{104}, 222405 (2014).

\bibitem{yang_memristive_2013}
Yang, J. J., Strukov, D. B. \& Stewart, D. R.
\newblock Memristive devices for computing.
\newblock {\em Nature Nanotech.} \textbf{8}, 13--24 (2013).

\bibitem{sengupta_proposal_2016}
Sengupta, A., Shim, Y. \& Roy, K.
\newblock Proposal for an all-spin artificial neural network: emulating neural and synaptic functionalities through domain wall motion in ferromagnets.
\newblock {\em IEEE Trans. Biomed. Circuits Syst.} \textbf{99}, 1--9 (2016).

\bibitem{bunyk_architectural_2014}
Bunyk, P. I. \emph{et al}.
\newblock Architectural considerations in the design of a superconducting
  quantum annealing processor.
\newblock {\em IEEE Trans. Appl. Supercond.} \textbf{24}, 1--10 (2014).

\bibitem{lemieux_design_2004}
Lemieux, G. \& Lewis, D.
\newblock \emph{Design of interconnection networks for programmable logic.}
\newblock ({Springer, Boston}, 2004).

\end{thebibliography}

%TC:ignore
%TC:ignore
\section*{Methods}
\begin{figure*}[h]
%-1/4 - 10*(cos(x))*(cos(x)) + (4-1/4)/(sin(x)*sin(x))
%from x = 0.6 to 2.5
% 	  y = 1.5 3
\includegraphics[width=\textwidth]{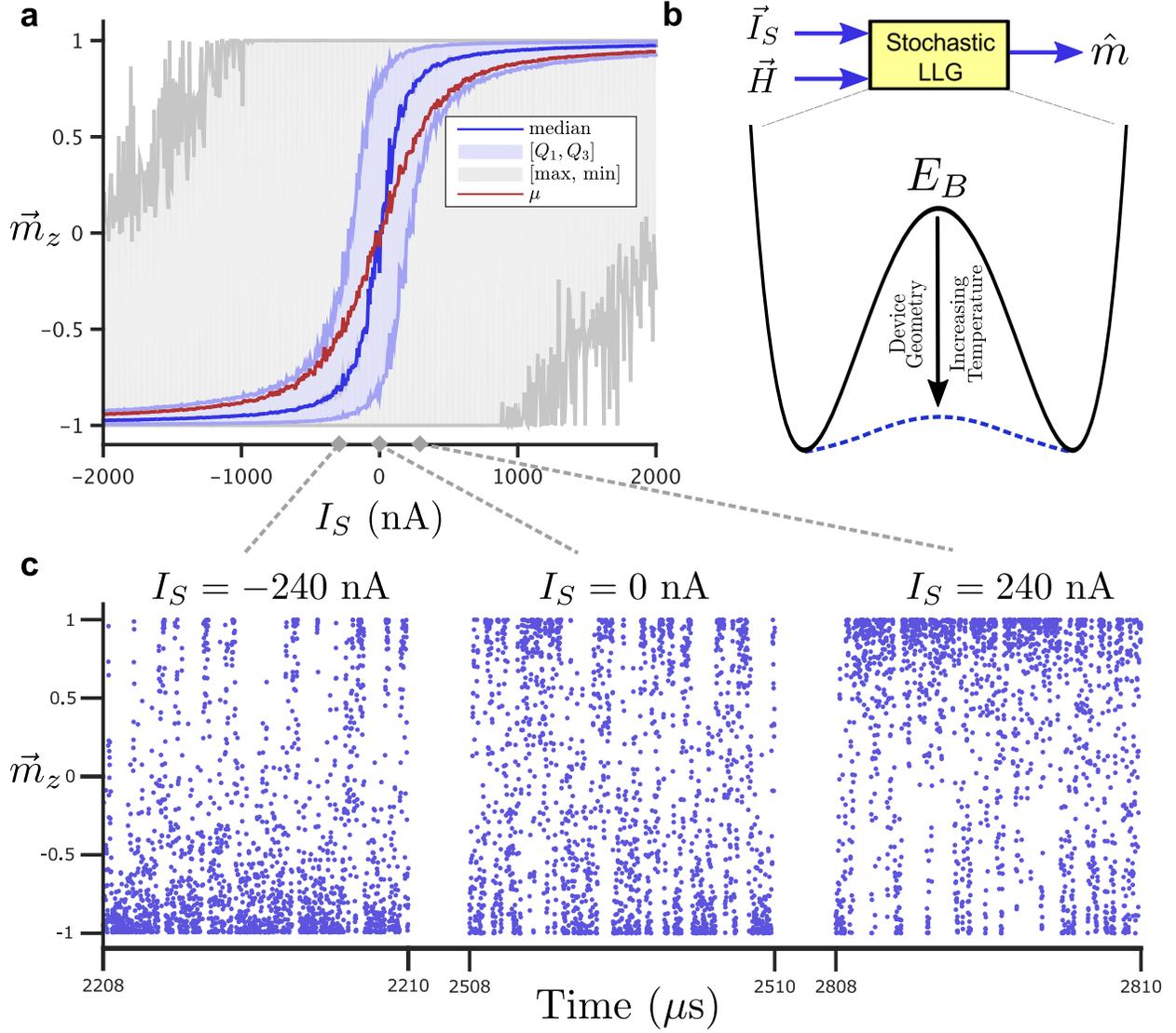}
    \phantomsubfigure{fig:fig1a}%
    \phantomsubfigure{fig:fig1b}%
    \phantomsubfigure{fig:fig1c}%
\caption{\label{fig:fig1}{\bf Response of stochastic nanomagnet to spin current.} {\bf a,} The magnetization of a stochastic nanomagnet is shown for varying spin currents. The five number summary of the magnetization $\vec{m}_z$ is shown throughout the simulation. {\bf b,} Obtaining stochastic operation for a magnet can be accomplished with a reduction of the energy barrier of the magnet $E_B$ through device geometry or by increasing its temperature. The response of the magnet to thermal noise under these conditions is modeled using a stochastic Landau-Lifshitz-Gilbert (LLG) circuit element based on the input spin current $I_S$ and magnetic field $H$. {\bf c,} Sample time slices are shown at various set points along the sigmoid in order to visualize the magnetization dynamics.}
\end{figure*}

\begin{figure*}[h]
	
\includegraphics[width=0.7\textwidth]{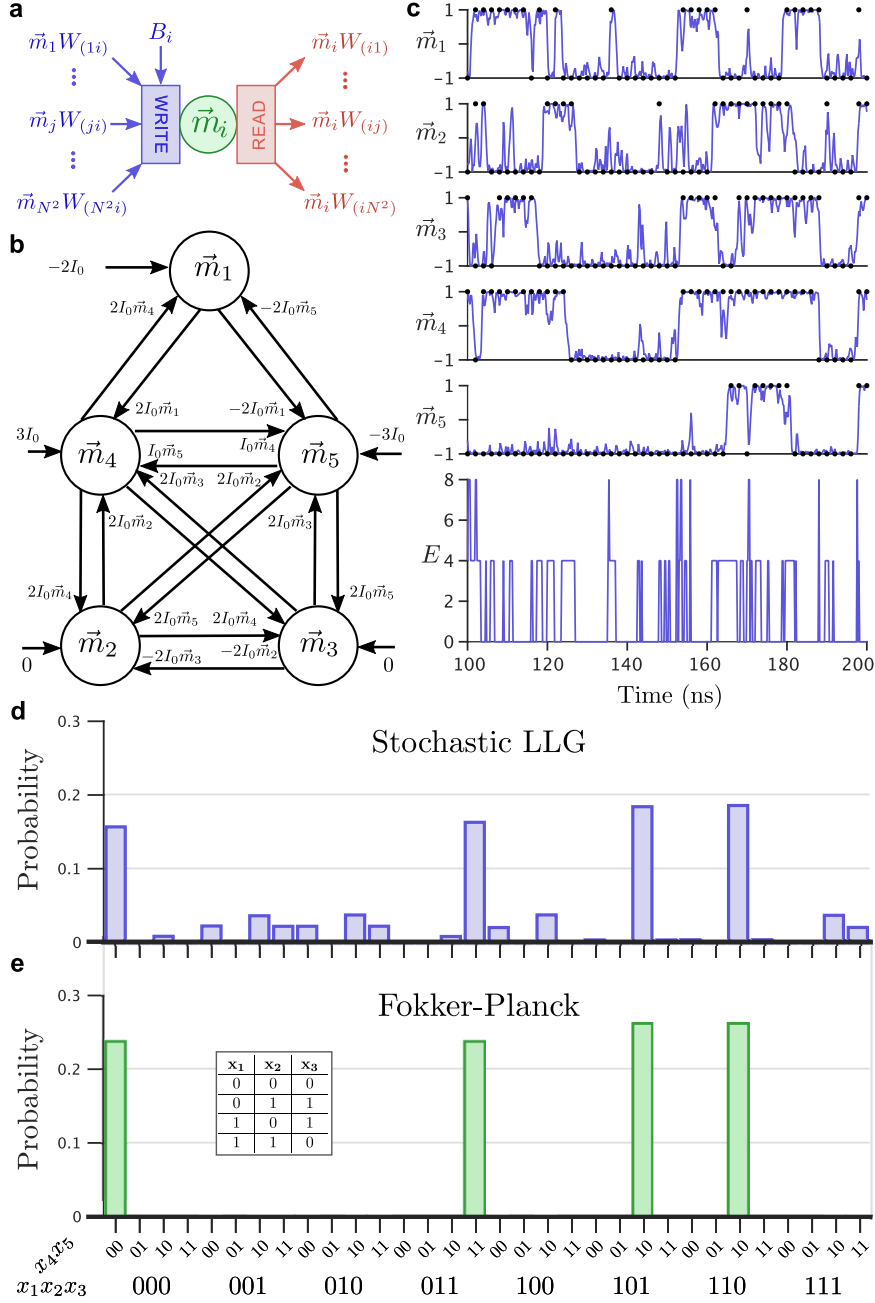}
    \phantomsubfigure{fig:fig2a}%
    \phantomsubfigure{fig:fig2b}%
    \phantomsubfigure{fig:fig2c}%
    \phantomsubfigure{fig:fig2d}%
    \phantomsubfigure{fig:fig2e}%
\caption{\label{fig:fig2}{\bf Boolean satisfiability with stochastic magnets.} {\bf a,} The coupling between individual magnets is represented using abstract write and read units. Each magnet $\vec{m}_i$ is influenced by problem specific on-site bias, $B_i$, and weighted, $W_{(ji)}$, coupling to magnet $\vec{m}_j$. In turn, magnet $\vec{m}_i$ influences magnet $\vec{m}_j$ through weight $W_{(ij)}$. {\bf b,} The truth table of the exclusive or operation is found by decomposing the operation into an expression involving only NOT, OR, and AND. The energy of this expression is found using the Ising model with two additional ancilla bits. The logical bits are represented by magnets $\vec{m}_i$ that are coupled and biased as specified by the Ising energy expression. {\bf c,} The magnetization response of the five magnets during a small time slice is digitized in order to compute the overall energy of the system as a function of time {\bf d,} Each digitized magnetization is used to represent the logical bits $x_i$. {\bf e,} Steady-State Fokker-Planck equation analytical solution using thresholding, demonstrating a qualitative match to the stochastic LLG solution of {\bf d}.}
\end{figure*}

\begin{figure*}[h]
%-1/4 - 10*(cos(x))*(cos(x)) + (4-1/4)/(sin(x)*sin(x))
%from x = 0.6 to 2.5
% 	  y = 1.5 3
\includegraphics[width=\textwidth]{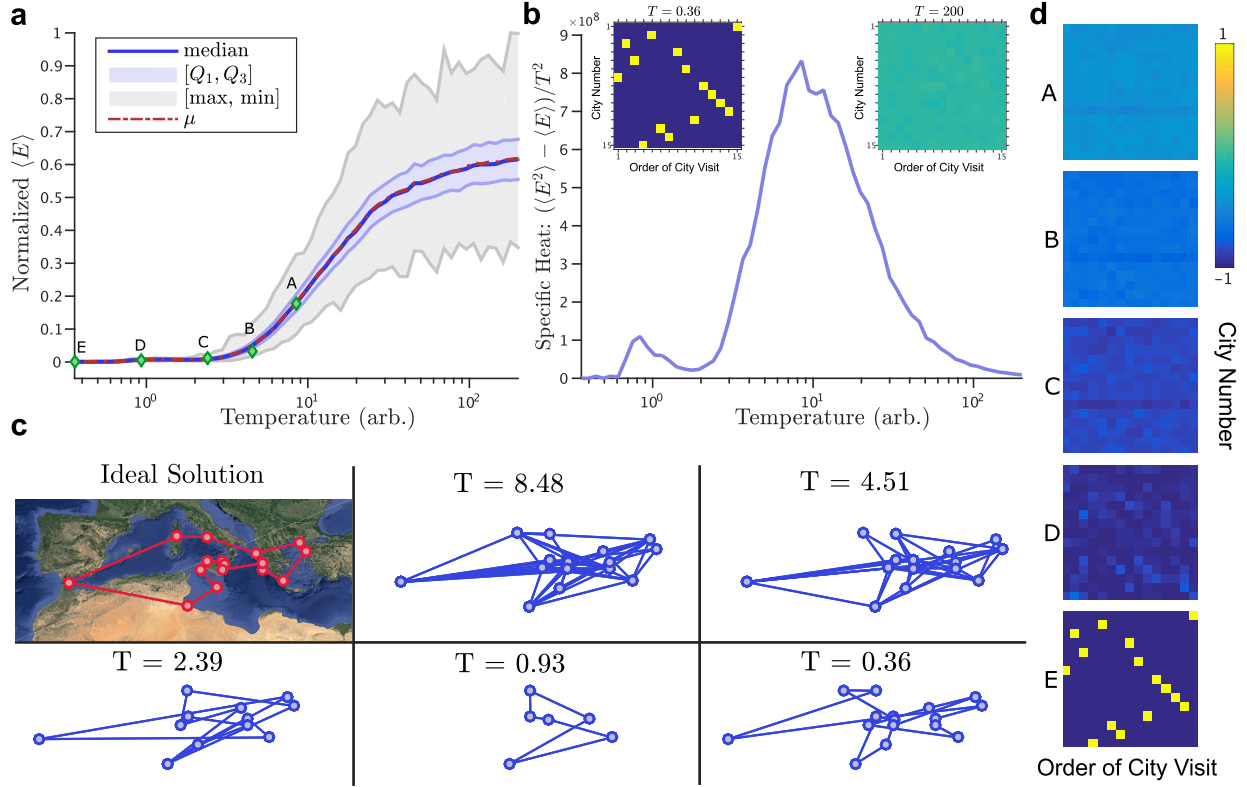}
    \phantomsubfigure{fig:fig3a}%
    \phantomsubfigure{fig:fig3b}%
    \phantomsubfigure{fig:fig3c}%
    \phantomsubfigure{fig:fig3d}%
\caption{\label{fig:fig3}{\bf Annealing stochastic nanomagnets for heuristic optimization of the traveling salesman problem.} {\bf a,} An $N = 16$ city traveling salesman problem based on the ulyssess16 data set\cite{reinelt_tsplib--_1991} was simulated using an array of $(N-1)^2 = 225$ stochastic magnets, assuming a fixed starting city. Each magnet represents if city $i$ was stop $j$ using $m_z = +1$ or was skipped $m_z = -1$ (insets). The magnets are prepared in a random initial configuration and gradually annealed until eventually frozen in a low-energy configuration. The normalized average energy of the system at each temperature is shown as the system is gradually annealed. {\bf b,} The specific heat of the array versus temperature is shown along with insets of the array configuration at early and late temperatures.  {\bf c,} The state of the array is shown as a TSP graph at various temperatures, shown as green diamonds in {\bf a}, during the annealing process with the ideal configuration shown on the top left. {\bf d,} Average magnetization shown at the temperatures of {\bf c}. (Map imagery data: Google, TerraMetrics)}
\end{figure*}

\noindent Simulations based on the modular framework for spintronics\cite{camsari2015modular} were used to produce the results in this work. Within the framework, a stochastic Landau-Lifshitz-Gilbert (LLG) model was used to simulate each nanomagnet. The magnetic parameters for the telegraphic PMA magnets used in the simulations are: effective anisotropy of PMA, $H_K^{\rm eff}=600 \ \rm Oe$,  saturation magnetization, $M_s=300 \ \rm emu/cc$, damping coefficient, $\alpha=0.01$, and PMA diameter, $\Phi=45 \ \rm nm$, amounting to a barrier height of $\Delta=1 \ \rm kT$. In all simulations, the initial state of the magnetic array was randomly selected. Figure \ref{fig:fig1} was produced using a modular stochastic LLG simulation element with the input current swept from $-2 \mu$A to $2 \mu$A in increments of 800 nA. At each current, the response of the magnet is observed for 10 $\mu$s. Figure \ref{fig:fig2} was simulated for 100 $\mu$s using the coupling depicted in the Figure and a reference current $I_0$ of $2 \mu$A. Figure \ref{fig:fig3} used an annealing schedule of $T_{i+1} = 0.9 T_i$ and Lagrange multiplier of $\lambda = 0.9/\text{max}(W_{(uv)})$. At each temperature the magnets were allowed to randomly walk for 1 $\mu$s and were measured every $200$ ps. The SAT and TSP magnetic networks were simulated using coupled stochastic LLG models with the intermagnet-coupling and on-site biases produced via the spin current term of the LLG equation. The magnetization of each magnet was digitized using Schmitt Trigger based thresholds. HSPICE was used to solve the simultaneous coupled differential equations of the magnetic network.

\begin{acknowledgments}
\noindent This work was supported in part by C-SPIN, one of six centers of STARnet, a Semiconductor Research Corporation program, sponsored by MARCO and DARPA and in part by the National Science Foundation through the NCN-NEEDS program, contract 1227020-EEC.
\end{acknowledgments}

\newpage

\clearpage
\newpage
\begin{table}
\centering
\begin{tabular}{c | l}
\hline
{\bf Function} & {\bf Technique} \\
\hline
\multirow{2}{*}{{\bf Writing}}&Spin-Orbit Torque \cite{liu_spin-torque_2012, sanchez2013spin}\\
&Voltage Control \cite{heron2014deterministic}\\
\hline
\multirow{2}{*}{{\bf Reading}}&Spin-Valves/Tunnel Junctions \cite{parkin_magnetically_2003} \\
&Inverse Spin-Hall Effect \cite{liu_spin-torque_2012}\\
\hline
\multirow{2}{*}{{\bf Amplification}}& Spin-Switches \cite{datta2012non}\\
&CMOS \cite{tangel_cmos_2004} \\
\hline
\multirow{4}{*}{{\bf Weighting}} & Floating-gate Regulators \cite{diep_spin_2014} \\
& Memristive Elements \cite{locatelli_spin-torque_2014,yang_memristive_2013,sengupta_spin_2015}\\
& Digital Logic \cite{yamaoka_20k-spin_2016}\\
& Fixed Voltages \\
\hline
\multirow{2}{*}{{\bf Routing}} & Tailored Topologies \cite{bunyk_architectural_2014}\\
& FPGA-Like Interconnect \cite{lemieux_design_2004}\\
\hline
\end{tabular}
\caption{\label{tab:phys} {\bf Options for Physical Realization:} Many options exist for physical realization of the proposed system of stochastic nanomagnets. These magnets must be written, read, possibly amplified, weighted, and routed for the network to form a Boltzmann machine. The design options shown in this table reflect various approaches that can be used to perform each of these functions.}
\end{table}

%TC:ignore

\end{document}